\documentclass[11pt]{article}\textheight 23cm
\usepackage[T1]{fontenc}
\usepackage[sc]{mathpazo}
\usepackage[utf8]{inputenc}
\usepackage[margin=1in]{geometry}
\usepackage{tikz}
\usepackage{graphicx}
\usepackage{amsmath}
\usepackage[export]{adjustbox}
\usepackage{titlesec}
\usepackage{epstopdf}
\usepackage{caption}
\usepackage{placeins}
\usepackage[colorlinks]{hyperref}
\hypersetup{colorlinks,breaklinks,
            urlcolor=[rgb]{0,0.5,0.5},
            linkcolor=[rgb]{0,0.5,0.5}}
\usepackage{color}
\titlelabel{\thetitle.\quad}
\usepackage{cuted}
\usepackage{tikz}
\usetikzlibrary{positioning}
\usepackage[position=top]{subfig}
\usepackage{verbatim}
\usepackage{caption}
\usepackage[toc,page]{appendix}
\usepackage[numbers,sort&compress]{natbib}
\begin{document}
\begin{center}
{
\huge Euler Number and Percolation Threshold on a Square Lattice with Diagonal Connection Probability and Revisiting the Island-Mainland Transition}\\ \vskip 0.5cm
Sanchayan Dutta$^1$, Sugata Sen$^2$, Tajkera Khatun$^3$, Tapati Dutta$^4$ and Sujata Tarafdar$^{5*}$\\
\vskip 0.5cm

$^1$ \footnotesize{Department of Electronics and Telecommunication, Jadavpur University, Kolkata 700032, India}\\
$^2$ \footnotesize{Department of Electrical Engineering, Jadavpur University, Kolkata 700032, India}\\
$^3$ \footnotesize{Department of Physics, Charuchandra College, Kolkata 700029, India}\\
$^4$ \footnotesize{Department of Physics, St. Xavier's College, Kolkata 700016, India}\\
$^5$ \footnotesize{Condensed Matter Physics Research Centre, Department of Physics,\\ Jadavpur University, Kolkata 700032, India}\\
$^*$ Corresponding Author. Email: \url{sujata_tarafdar@hotmail.com}.\\ Phone: +913324146666 (Extn. 2760), Fax: +913324138917
\end{center}
\vskip .5cm
\noindent {\bf \large{Abstract}}\\
\footnotesize{We report some novel properties of a  square lattice filled with white sites, randomly occupied by black sites (with probability $p$). We consider connections up to second nearest neighbours, according to the following rule. Edge-sharing  sites, i.e. nearest neighbours of similar type are always considered to belong to the same cluster. A pair of black corner-sharing sites, i.e. second nearest neighbours may form a 'cross-connection' with a pair of white corner-sharing sites. In this case assigning \textit{connected} status to both pairs simultaneously, makes the system quasi-three dimensional, with intertwined black and white clusters. The two-dimensional character of the system is preserved by considering the black diagonal pair to be \textit{connected} with a probability $q$, in which case the crossing white pair of sites are deemed disjoint. If the black pair is disjoint, the white pair is considered connected. In this scenario we investigate (i) the variation of the Euler number $\chi(p) \ [=N_B(p)-N_W(p)]$ versus $p$ graph for varying $q$, (ii) variation of the site percolation threshold with $q$ and (iii) size distribution of the black clusters for varying $p$, when $q=0.5$. Here $N_B$ is the number of black clusters and $N_W$ is the number of white clusters, at a certain probability $p$. We also discuss the earlier proposed 'Island-Mainland' transition (Khatun, T., Dutta, T. \& Tarafdar, S. Eur. Phys. J. B (2017) 90: 213) and show mathematically that the proposed transition is not, in fact, a critical phase transition and does not survive finite size scaling. It is also explained mathematically why clusters of size 1 are always the most numerous.}\\

\section{Introduction}
\label{Introduction}
Different aspects of the properties of two-dimensional square lattices has been an ongoing challenge for over half a century. Yet, there are certain lattice properties which have not been as well studied as the others. 

The identification of the percolation transition as a critical phase transition has been a significant finding with deep theoretical as well as practical implications \cite{stauffer}.
Another quantity which survives finite size scaling is the \textit{Euler number} which has therefore many practical applications. The concept of \textit{Euler number} is an important topological property inspired from ideas useful to the field of image processing \cite{dey2007co}.The Euler number (or $genus$) is defined as the difference between the number of "connected components" and the number of "holes" in an image. These type of topological properties remain invariant under any arbitrary \textit{rubber-sheet} transformation, i.e. stretching, shrinking, rotation etc. and thus are very useful in image characterization to match shapes, recognize objects, image database retrieval and other image processing and computer vision applications. Analysis of images of real systems like soil crack patterns \cite{vogel2005studies, Khatun2017}, fast reading of car number plates \cite{Faqheri2009} and automatic signature matching \cite{Vatsa2004} have been facilitated through use of Euler numbers.
In diagnostic imaging, analysis of patterns with proper thresholding, is extremely important to identify irregularities indicating possible medical conditions.  Here again the  Euler number plays an important role \cite{Wong2007,Zhang2006} 

Recently the Euler number and its variation with site occupation probability  on a square lattice, has been discussed by Khatun \textit{et al.} \cite{Khatun2017}. Black ($B$) unit squares are randomly dropped,  with probability $p$ onto a lattice initially filled with white ($W$) unit squares. Here sites up to second nearest neighbours are considered to be connected. That is, by definition edge sharing as well as corner sharing sites of similar type belong to the same cluster. A problem in this situation is that with clusters defined thus, there may appear points where two diagonal connections cross each other and the system no longer remains ideally two-dimensional \cite{Khatun2017, feng} , but has to be visualized as a quasi-three dimensional system. In the present study we report an extension of the work by Khatun \textit{et al} \cite{Khatun2017}, where this problem is circumvented. A new variable $q$ is introduced, which represents the \textit{probability} of a pair of diagonal $B$ sites being connected, in which case the pair of diagonal $W$ sites sharing the same corner will be necessarily considered disjoint. Now the \textit{flattened system} can be represented as a purely two-dimensional lattice. The site percolation threshold $p_c(q)$,over the whole range of  $q$ covering values from 0 to 1 are presented. The number of black clusters($N_B$) is now a function of $p$ and $q$, so is the number of white clusters ($N_W$). The Euler number is defined as $\chi(p,q)= N_B(p,q) - N_W(p,q)$ so `connected components' and `holes' imply here clusters of occupied (Black/White) or vacant (White/Black) sites respectively. Random deposition and clustering on square lattices with nearest neighbour as well as second nearest neighbour connections have been studied earlier, but \textit{probabilistic} connection between second neighbours (introduced through $q$) is a new concept, which retains simultaneously the two-dimensional as well as stochastic character of the system.

Apart from the percolation threshold $p_c$, i.e. the value of $p$ where the $B$ sites first form a system-spanning `infinite cluster' the structure and size-distribution of the \textit{finite clusters} are also of great interest and considerable work has been done for two-dimensional lattices with various patterns \cite{grimmett_1999,bollobas}. The cluster size distributions in the new scenario are studied and it is shown that their qualitative features do not vary significantly with $q$. In addition we show mathematically  that an `island-mainland' transition, conjectured by \cite{Khatun2017} from numerical simulations  cannot be a critical phase transition and may be observed in finite-sized systems only.

Mertens-Ziff (2016) \cite{ziff} and Sykes-Essam (1964) \cite{sykes_essam} have also worked on the Euler characteristic $\chi_c$ albeit they follow a slightly different definition which involves the concept of matching lattices. On a square lattice if nearest neighbours (NN), i.e. edge-sharing sites of same type are considered to be connected, the Euler characteristic is defined as
$$\chi_c(p) = 1/L^{2}(N_B(p) - N_{WM}(p))$$ where $N_B(p)$ is the number of clusters of $B$ sites on the primary lattice and $N_{WM}(p)$ is the number of $W$ clusters on the \textit{matching lattice} corresponding to the primary lattice.
The matching lattice of the primary square lattice is obtained by adding edges to each face of the primary lattice such that the boundary vertices of that face form a clique, namely a
fully connected graph. For the square lattice, this means that
we add the two diagonals to each face: the matching lattice
of the square lattice is the square lattice with next-nearest
neighbours. 

\begin{figure}
\centering
\label{Quasi3D}
\includegraphics[max size={0.5\textwidth}{\textheight}, trim =0.1cm 0.2cm 0cm 0.1cm, clip]{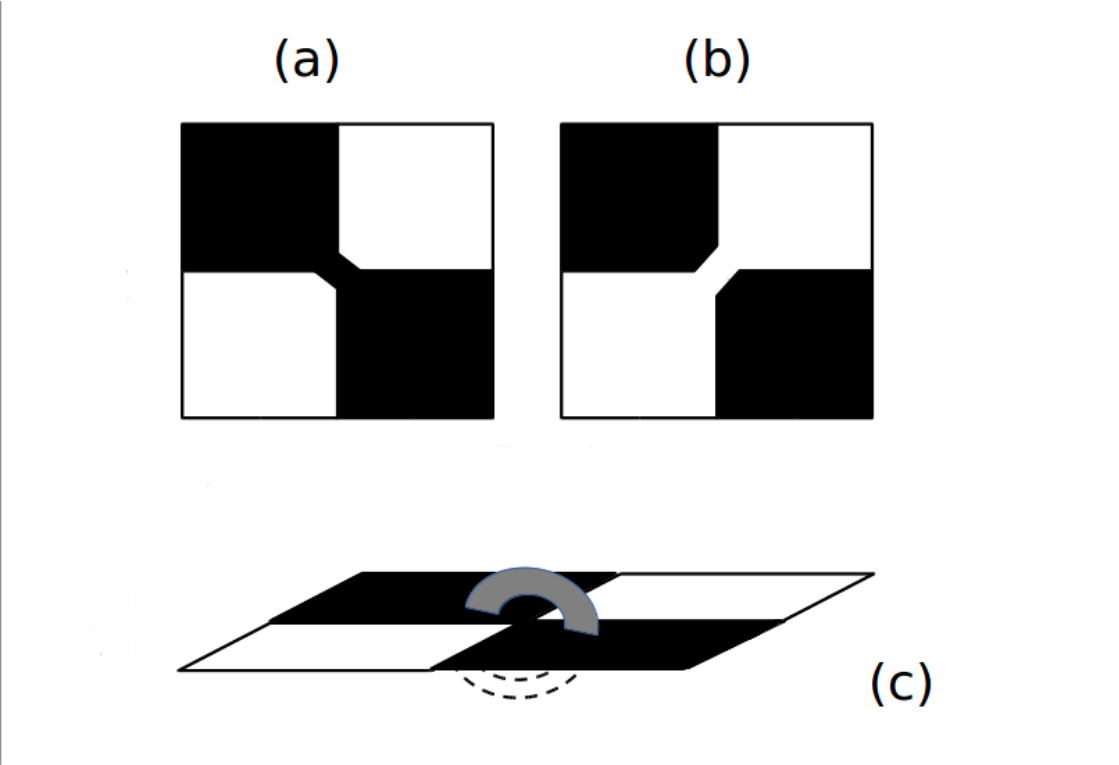}
\caption{Possible diagonal connections in a $2\times 2$ square lattice having two diagonally opposite black pixels (squares) and two diagonally opposite white pixels which orthogonally neighbour them. (a) shows a diagonal connection between two diagonal black pixels (squares) which disconnects the orthogonally neighbouring white pixels while (b) shows the reverse situation where two diagonal white squares are connected. In case both are connected, as in (c), the system becomes \textit{quasi}-3D.}
\end{figure}

Here we will focus on the first definition of Euler number, as defined in \cite{dey2007co} i.e. $\chi(p) = N_B(p) - N_{W}(p)$. This definition is equivalent to the case when the primary and complementary lattices are identical and connections of black and white clusters in the primary and complementary lattices are governed by the diagonal connection probability $q$ as described before.

The situation discussed here is connected to another practical problem of surface science, namely wetting, spreading or salt deposition on a plane surface. This depends on the properties of the spreading fluid and substrate (two different fluids may be involved to make things more complex). In case of crystal growth, for example, with a cubic crystal like NaCl crystallizing from a complex solution \cite{moutushi}, one may think of an underlying square lattice. Here, the crystal growth sometimes favours diagonal connections over edge connections. Crystal growth in this case is in the form of narrow fingers connected through corners, while in others it may grow as compact cubes or empty box-like hopper crystals.

We expect the present discussions to be applicable to wetting-spreading problems between fluids and substrates with complex interactions amongst themselves, in determining what final configurations the system shall take, since growth can happen either across the edge or the corner of a square lattice, but in a real situation will depend on the physics and chemistry governing the wetting or growth process.

Following this introduction, in the next section \ref{Sec 2} we present details of the numerical simulation and the results obtained are presented and discussed in section \ref{Sec 3}. In section \ref{Sec 4} we discuss the idea behind the Island-Mainland transition suggested in \cite{Khatun2017}, its limitations and its relationship with our model. Finally, section \ref{Sec 5} gives a discussion of the results and concludes with directions for future work.

\section{Simulation Details}
\label{Sec 2}

For our simulations all binary random matrices were generated using the Xorshift pseudo-random generator \cite{Xorshift} with system size as seed.

\subsection{Euler Number Variation with Diagonal Connection Probability $q$}\label{Sec 2.1}

\begin{figure}
\begin{tikzpicture}
\centering
\node (img) {\includegraphics[width=\textwidth, trim=0cm 0cm 0cm 0cm, clip]{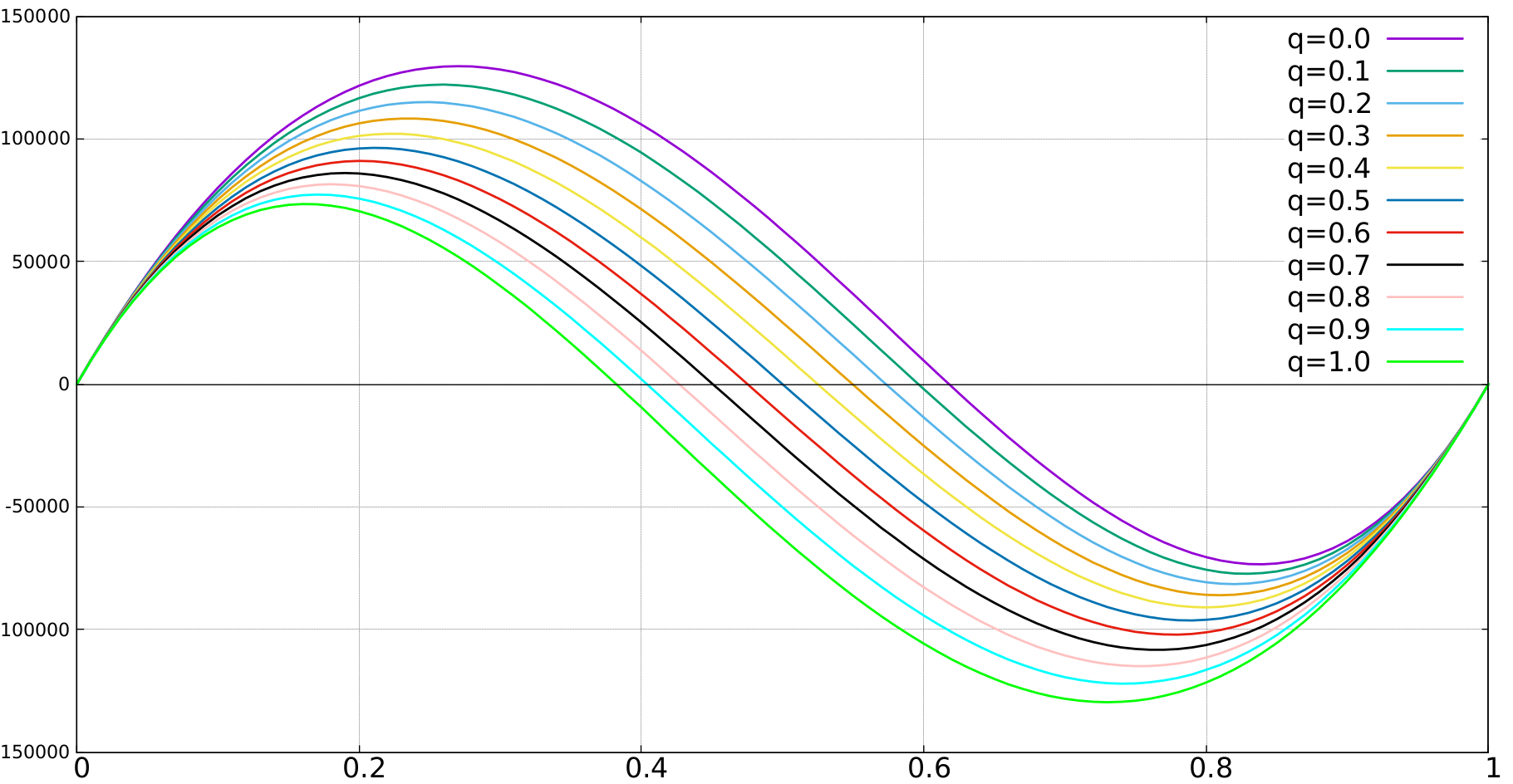}};
\node[below=of img, node distance=0cm, yshift=1cm,font=\color{black}] {$\mathrm{\mathbf{p}}$};
\node[left=of img, node distance=0cm, rotate=90, anchor=center,yshift=-0.7cm,font=\color{black}] {$\chi(\mathbf{p})$};
\end{tikzpicture}
\caption{Variation of Euler number $\chi(\mathbf{p})$ with site occupation probability $\mathbf{p}$ for different values of diagonal connection probabilities $\mathrm{q}$.}
\label{EulerNumber}
\end{figure}

Random binary matrices of size $1000\times 1000$ were generated for different values of occupation probability $p$ in the range $[0,1]$ in steps of $0.1$. A diagonal connection probability $q$ as described in section (\ref{Introduction}) is also taken into account. Clustering, with diagonal connection probability $q$ taken into account, was done dynamically during the process of generation of the random matrices, to avoid extra re-iterations through the whole lattice. Statistics for $\chi(p)$ were collected and averaged over for $100$ random matrices for each such value of $p$. The results have been plotted in figure \ref{EulerNumber}.

\begin{figure}
\centering
\begin{tikzpicture}
\node (img){\includegraphics[max size={0.7\textwidth}, trim=0.5cm 0.5cm 0cm 0cm, clip]{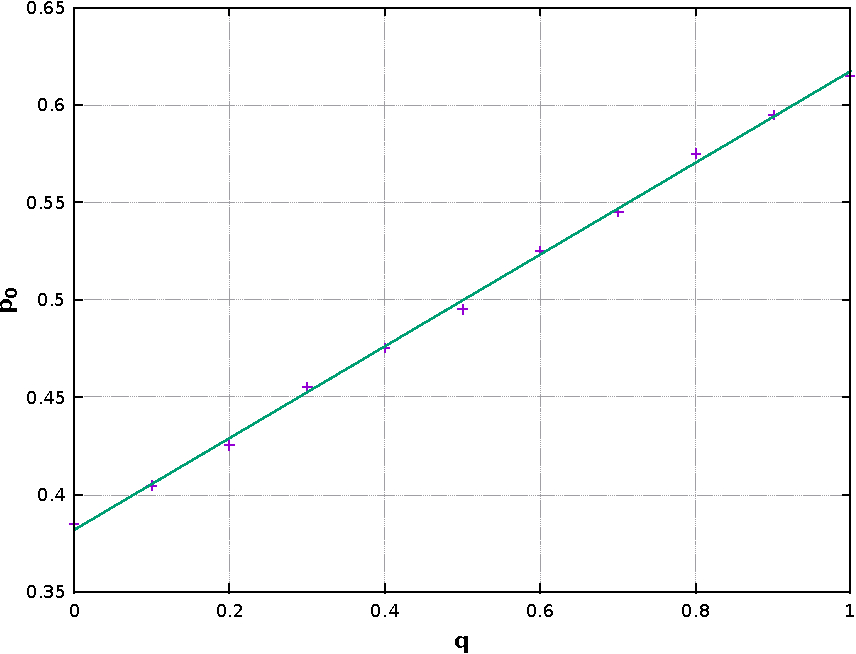}};
\node[left=of img, node distance=0cm, rotate=0, anchor=center,yshift=0.5cm,xshift=0.8cm, font=\color{black}] {$\mathbf{p}_{0}$};
\node[below=of img, node distance=0cm, xshift=0cm, yshift=1.3cm,font=\color{black}] {$\mathbf{q}$};
\end{tikzpicture}
\caption{A plot showing how the probability at which there exist an equal number of black and white clusters i.e. $\mathbf{p_0}$ varies with diagonal connection probability $\mathbf{q}$. The error bounds lie within the symbol size.}
\label{variation}
\end{figure}

Let us call the probability at which the curves for different values of $q$ cross the horizontal axis $p_0$ (which is a function of $q$). The variation of $p_0$ with $q$ is shown in figure \ref{variation} along with the regression line in blue. 

\subsection{Variation of Spanning Cluster Percolation Threshold with Diagonal Connection Probability $q$}\label{Sec 2.2}

Let $\Pi(p,L)$ be the probability that a square lattice of size $L\times L$ percolates at concentration $p$. We use the notion of \textit{site percolation} \cite{grimmett_1999, stauffer} here i.e. for some value of $p$ a path begins to exist between any two opposite pair of edges of the square lattice. In an infinite system we have $\Pi=1$ above $p_c$ and $\Pi=0$ below $p_c$. For finite systems $\Pi$ is expressed as $\Phi[(p-p_c)L^{1/\nu}]$
where $\nu$ is a critical exponent (which is zero for infinite systems). $\Phi$ is a monotonically increasing \textit{scaling function} which maps values in $(-\infty,\infty)$ to $(0,1)$. Since $\Pi$ is expected to approach the step function when $L\to\infty$, we might define an effective threshold at the concentration where $\Pi = 1/2$. This effective threshold $p_{\text{eff}}$ approaches the true percolation threshold $p_c$ when $L\to\infty$.  

The $p_{\text{eff}}$'s were first determined using a binary search approach. The two intial bounds  for $p$ were taken as $0.3$ and $0. 7$. We then iteratively checked for the particular value of p for which percolation probability $\Pi$ first hit $50\%$. For each value of $p$ considered during the iterations, the value of $\Pi$ was determined by averaging over 500 randomly generated square lattices (corresponding to the specific value of p). Three decimal places of accuracy was considered. The reason for choosing $0.3$ and $0.7$ was that, for all the system sizes and all values of $q$, $\Pi(p=0.3)$ was always $0$ and $\Pi(p=0.7)$ was always $1$. Thus, the percolation threshold had to lie within $0.3$ and $0.7$ and wouldn’t be outside that range in any case. The values were re-checked using the Monte Carlo method described in \cite[p.~73]{stauffer} upto the third decimal place.

\begin{figure}
\centering
\begin{tikzpicture}
\node (img){\includegraphics[max size={0.7\textwidth}, keepaspectratio, trim=0.61cm 0.5cm 0cm 0cm, clip]{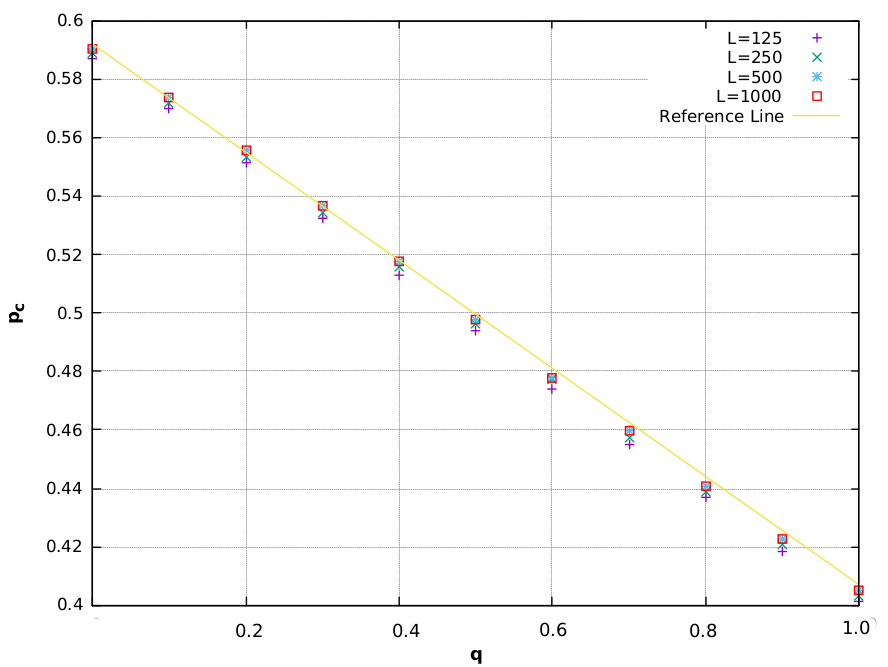}};
\node[left=of img, node distance=0cm, rotate=0, anchor=center,yshift=0cm,xshift=0.8cm, font=\color{black}] {$\mathbf{p}_{\text{eff}}$};
\node[below=of img, node distance=0cm, xshift=0cm, yshift=1.3cm,font=\color{black}] {$\mathbf{q}$};
\end{tikzpicture}
\caption{Percolation thresholds for different values of diagonal connection probabilities $\mathbf{q}$ and system sizes $\mathbf{L}$. The error bound for the data points is $\pm 0.005$.}
\label{ThresholdQ}
\end{figure}

We studied the variation of $p_{\text{eff}}$ for different values of $q$ and $L$. To be more specific, we calculated $p_{\text{eff}}$ by averaging over $500$ randomly generated binary matrix configurations of sizes $L=125, 250, 500$ and $1000$ each, with $q$ varying from $0$ to $1$, in steps of $0.1$. The results have been plotted in Figure \ref{ThresholdQ}. The ``Reference Line'' in the figure is the line which passes through the coordinates $(0,0.592)$ and $(1,0.407)$ and corresponds to $L\to\infty$ percolation thresholds. The boundary point coordinates of the reference were obtained from the 2005 paper by Malarz and Galam \cite{malarz_galam}. In between these two boundary points the functional form of the percolation threshold $p_c$ is 
$$p_c = p_\text{eff} (L\to\infty) =  -0.185 q + 0.592.$$ 

When considering only the Von Neumann
($\mathrm{N}^2$)\footnote{The neighborhood composed of a central cell and its four adjacent cells, on a two-dimensional square lattice.} neighborhood the site percolation threshold is approximately $0.592$ and when considering the Moore ($\mathrm{N}^2+\mathrm{N}^3$)\footnote{The neighborhood composed of a central cell and the eight cells which surround it, on a two-dimensional square lattice.} neighborhood  the site percolation threshold is approximately $0.407$. The first case essentially corresponds to the $q=0$ case and the second case corresponds to the $q=1$ case.

\begin{figure}
\captionsetup[subfigure]{labelformat=empty}
\centering
\begin{tikzpicture}
\node (img){\subfloat[]{\label{main:c}\includegraphics[width=\textwidth, trim=0cm 0cm 0cm 0cm, clip]{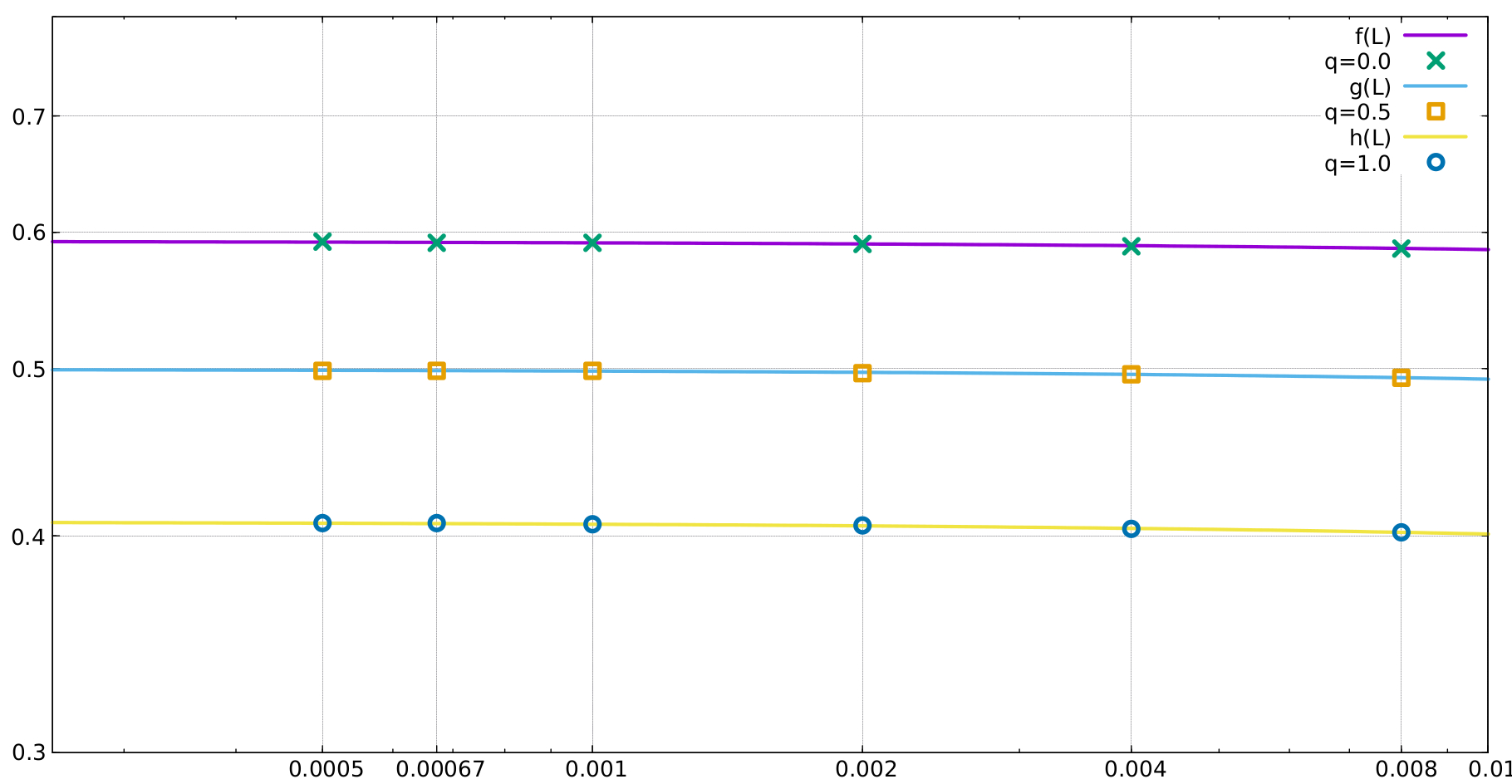}}};
\node[left=of img, node distance=0cm, rotate=0, anchor=center,yshift=0cm,xshift=0.4cm, font=\color{black}] {$\mathbf{p}_{\text{eff}}$};
\node[below=of img, node distance=0cm, xshift=0cm, yshift=1.3cm,font=\color{black}] {$1/\mathbf{L}$};
\end{tikzpicture}
\caption{Finite-size scaling using power-law fit, for $q=0$, $q=0.5$ and $q=1$ respectively. Data was collected for $\mathbf{L}=125, 250, 500, 1000, 1500$ and $2000$, $p_{\text{eff}}$ is plotted against $1000/\mathbf{L}$ in log-log scale. Resulting graphs are linear, the equations are given in the text. The error bounds lie within the symbol sizes.}
\label{FScalingPlot}
\end{figure}

Furthermore, we used the method of finite size scaling to estimate the actual percolation thresholds $p_c$ for different values of $q$. We know that $|p_{\text{eff}}(L) - p_c| \propto L^{-\frac{1}{\nu}}$ where $\nu$ is a percolation critical exponent which has a standard value of $\frac{4}{3}$ for dimension $d=2$ lattices. According to the universality principle the value of the critical exponents are independent of local details \cite{stauffer} as they describe the system in the limit where the correlation length diverges. We performed a power law fit on the $(1/L)$ vs. $p_{\text{eff}}$ data (for different values of $q$), obtaining the predicted values of the percolation thresholds as well as the value of $\nu \approx \frac{4}{3}$, that is, the obtained values of $\frac{1}{\nu}$ from equations (\ref{a}), (\ref{b}) and (\ref{c}) turn out to be close to the expected value of $\frac{3}{4}$ (for $d=2$ lattices). In figure \ref{FScalingPlot} the power law fit has been shown for $q=0$, $q=0.5$ and $q=1$ respectively, in a double log scale. The best fit equations for the three values of $q$, as shown in figure \ref{FScalingPlot} are  as follows:
for $q=0$
\begin{equation}p_{\text{eff}}(L) = f(L)=0.59299-0.13493L^{-0.64809} \tag{a}\label{a}\end{equation}
for $q=0.5$
\begin{equation}p_{\text{eff}}(L) = g(L)=0.49994-0.15253L^{-0.67654}\tag{b}\label{b}\end{equation}
and for $q=1$
\begin{equation}p_{\text{eff}}(L) = h(L)=0.40799-0.13492L^{-0.64809}.\tag{c}\label{c}\end{equation}

\begin{figure}
\centering
\begin{tikzpicture}
\node(img){\includegraphics[width=0.7\textwidth]{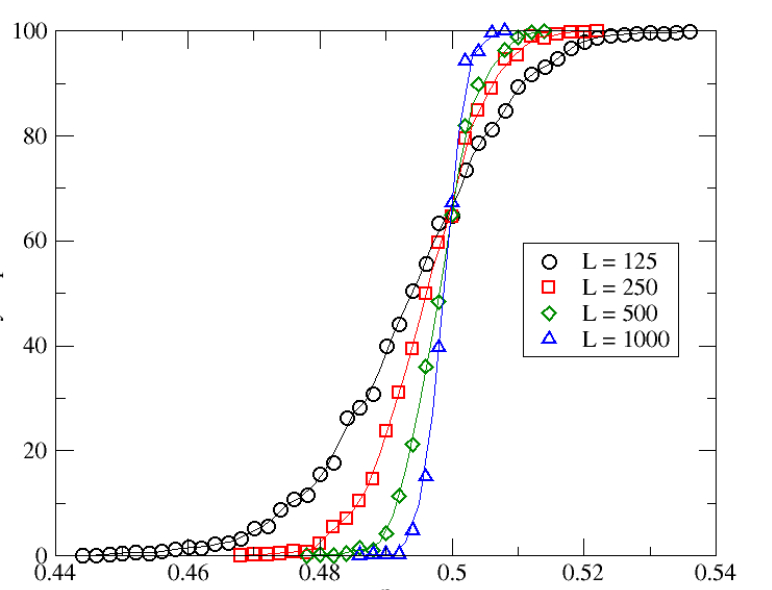}};
\node[left=of img, node distance=0cm, rotate=0, anchor=center,yshift=0.5cm,xshift=0.8cm, font=\color{black}] {$\Pi(\mathbf{p})$};
\node[below=of img, node distance=0cm, xshift=0cm, yshift=1cm,font=\color{black}] {$\mathbf{p}$};
\end{tikzpicture}
\caption{A plot showing the percentage percolation probability $\Pi$ for different values of site occupation probability $\mathbf{p}$, as obtained from our simulations. Results were averaged over $500$ iterations for each system size $\mathbf{L}$. The running average method was used to approximate the data points and estimate the critical percolation probability, which turns out to be $0.499$ when diagonal connection probability $q$ is $0.5$.}
\label{PiVariation}
\end{figure}

The variation of $\Pi$ with $p$ for different system sizes $L$, with $q$ fixed at $0.5$, is shown in Figure \ref{PiVariation}. The intersection of the system sizes indicates a value of $0.500$ for the percolation threshold with a percolation probability $64.6\%$.

\clearpage

\subsection{Size Distribution of Clusters}\label{sec 2.3}

\begin{figure}
\centering
\begin{tikzpicture}
\node (img){\includegraphics[scale=.7, trim=3cm 1cm 0cm 0cm, clip]{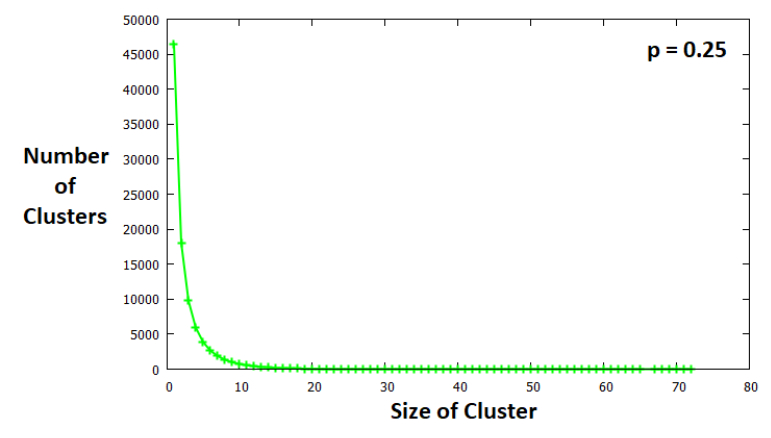}};
\node[left=of img, node distance=0cm, rotate=90, anchor=center,yshift=-1cm,xshift=0cm, font=\color{black}] {\scriptsize{\textbf{Number of Clusters}}};
\node[below=of img, node distance=0cm, xshift=0cm, yshift=1cm,font=\color{black}] {\scriptsize{\textbf{Size of Clusters}}};
\end{tikzpicture}
\caption{In the sub-critical phase when $p=0.25$ an exponential decay is observed.}
\label{BSD025}
\end{figure}

\begin{figure}
\centering
\begin{tikzpicture}
\node (img){\includegraphics[scale=0.5, trim=0cm 0cm 0cm 0cm, clip]{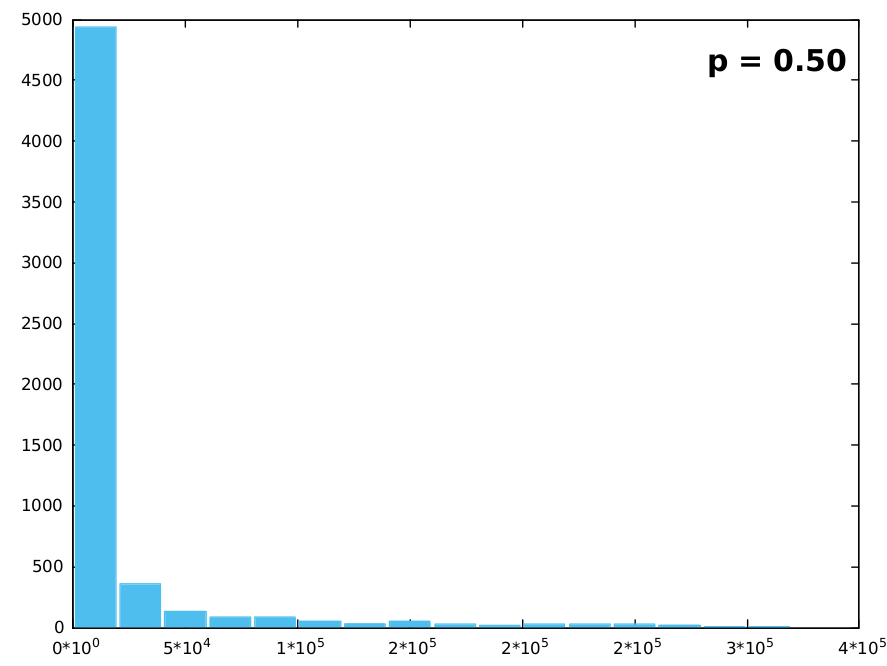}};
\node[left=of img, node distance=0cm, rotate=90, anchor=center,yshift=-0.7cm,xshift=0cm, font=\color{black}] {\scriptsize{\textbf{Number of Clusters}}};
\node[below=of img, node distance=0cm, xshift=0cm, yshift=1cm,font=\color{black}] {\scriptsize{\textbf{Size of Clusters}}};
\end{tikzpicture}
\caption{Nearby the critical phase i.e. $p = 0.5$, clusters are seen to vary over a wide range and hence nearby cluster sizes were binned together to observe the averaged statistics.}
\label{BSD050}
\end{figure}

\begin{figure}
\centering
\begin{tikzpicture}
\node (img){\includegraphics[scale=.5, trim=0cm 0cm 0cm 0cm, clip]{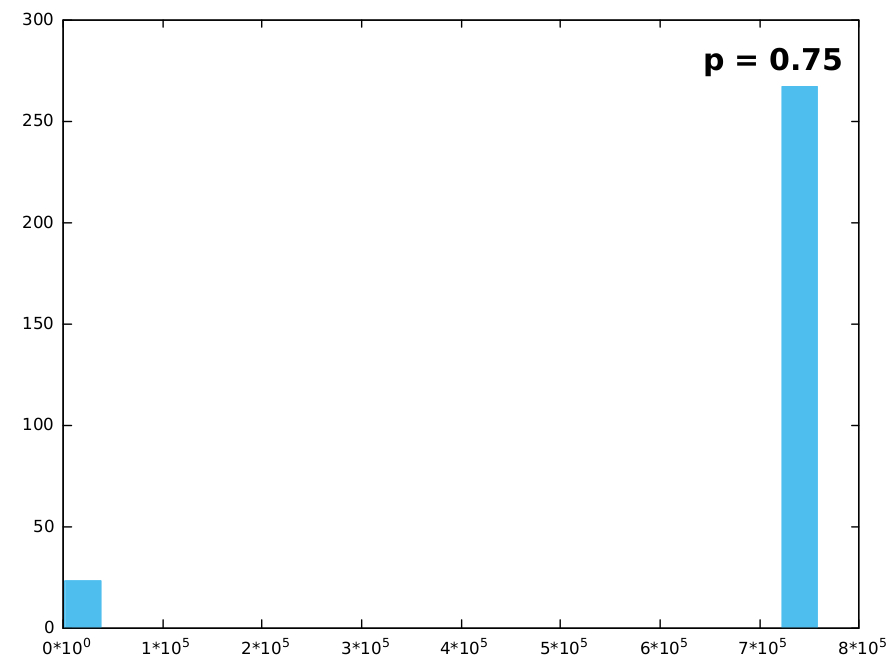}};
\node[left=of img, node distance=0cm, rotate=90, anchor=center,yshift=-1.1cm,xshift=0cm, font=\color{black}] {\scriptsize{\textbf{Number of Clusters}}};
\node[below=of img, node distance=0cm, xshift=0cm, yshift=1cm,font=\color{black}] {\scriptsize{\textbf{Size of Clusters}}};
\end{tikzpicture}
\caption{In the super-critical phase when $p=0.75$, only a single ``large" cluster was seen for each one of the randomly generated binary matrices and a small number of irregularly distributed small clusters.}
\label{BSD075}
\end{figure}

Cluster size statistics for $q=0.5$ are shown in figures \ref{BSD025}, \ref{BSD050} and \ref{BSD075}. Data were collected over $100$ randomly generated binary matrices with $p$ set at $0.25,0.5$ and $0.75$ respectively. The labelling and the subsequent counting of clusters was done using an extended version of the Hoshen-Kopelman algorithm \cite{hoshen_kopelman} which takes into account the diagonal connection probability $q$. 

For $p=0.25$ the size of $B$ clusters is confined to within 80 squares and the number of clusters of each size in the whole system is seen to fall exponentially.  As the occupation probability $p$ increases further cluster sizes increase by several orders of magnitude and it becomes necessary to bin the data into groups within certain ranges of magnitude. Data for $p=0.5$ and $p=0.75$ are shown thus in \ref{BSD025}, \ref{BSD050} and \ref{BSD075}. Statistics were collected and averaged over $100$ randomly generated binary matrices.

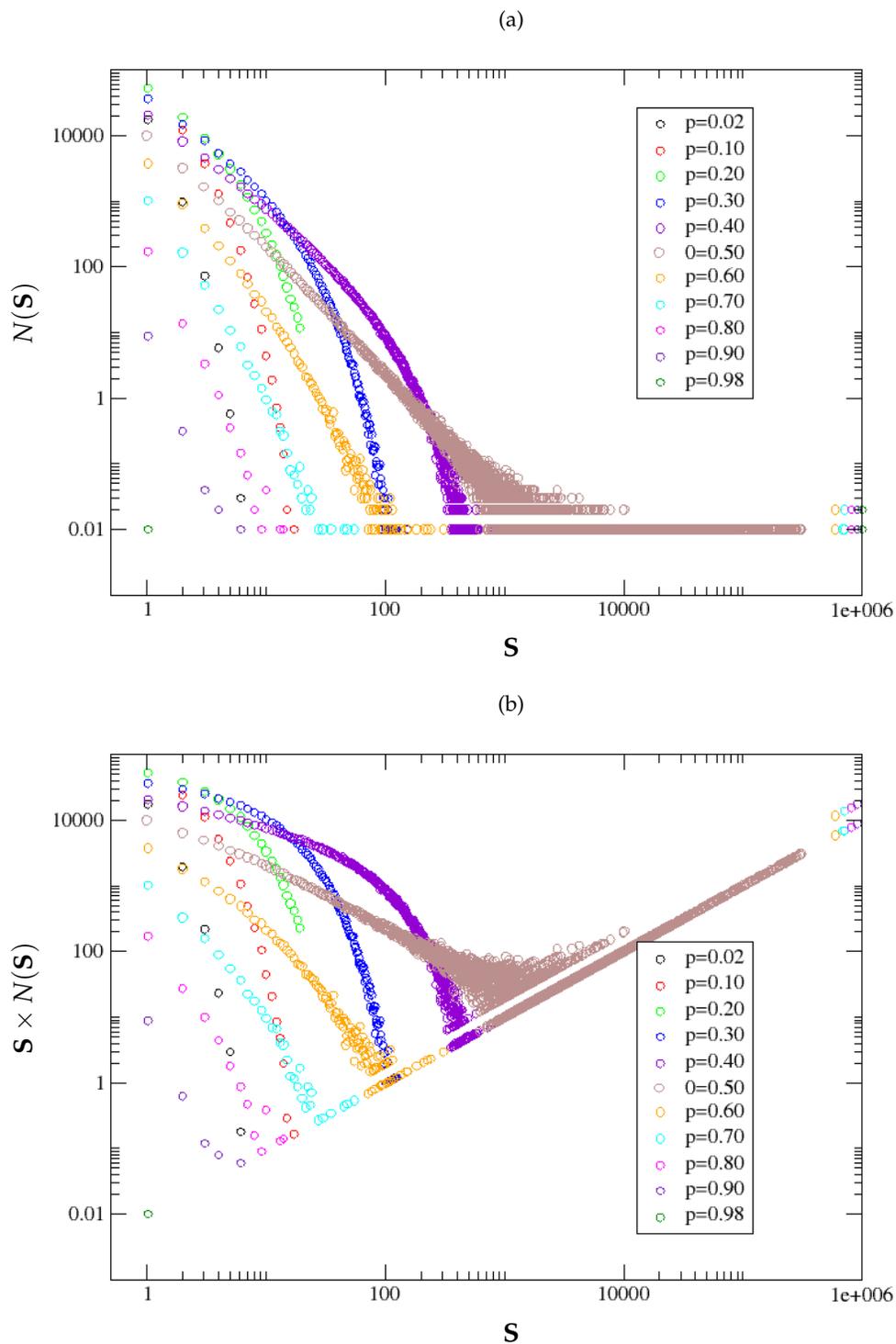
\begin{figure}
\centering
\begin{tikzpicture}
\node (img){\subfloat[]{\label{SN(S)}\includegraphics[width=0.8\textwidth, trim=1.3cm 2cm 0cm 2cm, clip]{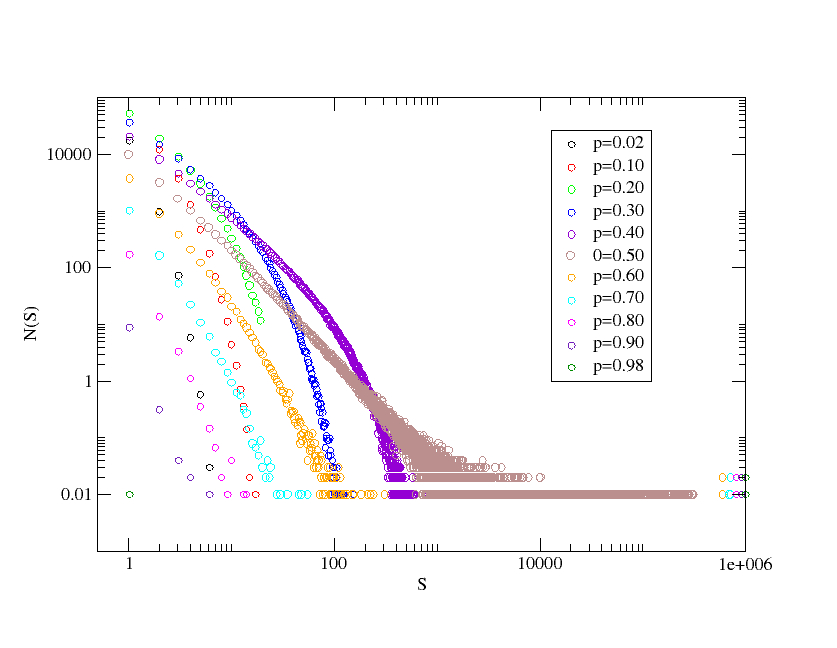}}};
\node[below=of img, node distance=0cm, xshift=0cm, yshift=1cm,font=\color{black}] {$\mathrm{\mathbf{S}}$};
\node[left=of img, node distance=0cm, rotate=90, anchor=center,xshift=0cm, yshift=-0.7cm,font=\color{black}] {$N(\mathbf{S})$};
\end{tikzpicture}
\par\medskip
\begin{minipage}{\linewidth}
\centering
\begin{tikzpicture}
\node (img){\subfloat[]{\label{N(S)}\includegraphics[width=0.8\textwidth, trim=1.3cm 2cm 0cm 2cm, clip]{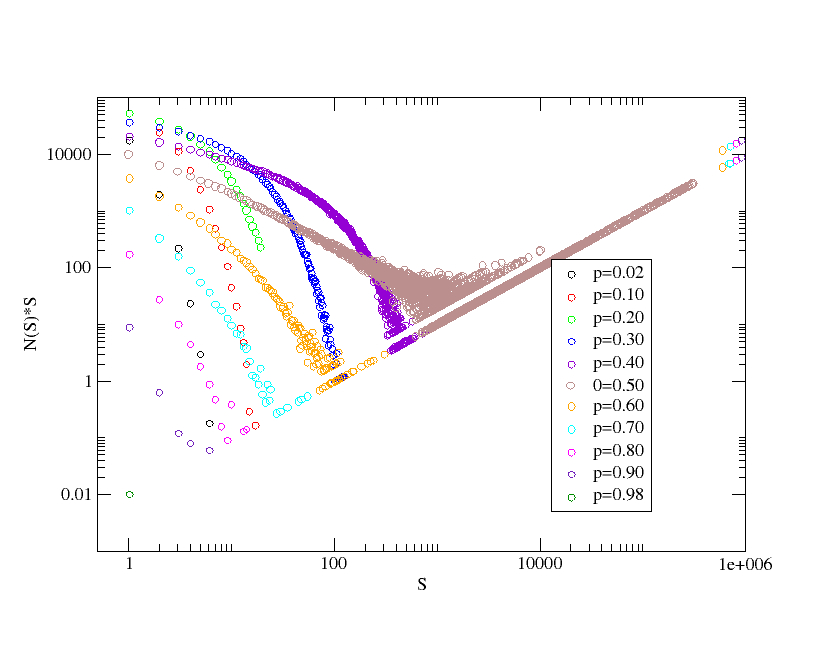}}};
\node[below=of img, node distance=0cm, xshift=0cm, yshift=1cm,font=\color{black}] {$\mathbf{S}$};
\node[left=of img, node distance=0cm, rotate=90, anchor=center,xshift=0cm, yshift=-0.7cm, font=\color{black}] {$\mathbf{S}\times N(\mathbf{S})$};
\end{tikzpicture}
\end{minipage}
\caption{(a) $\mathbf{S}$ represents cluster size. $N(\mathbf{S})$ represents the number of clusters of a certain size. Based on our simulations, we plot the nature of variation of $N(\mathbf{S})$  against cluster size $\mathbf{S}$ for different values of occupation probability $\mathrm{p}$ in double logarithmic scale. (b) $N(\mathbf{S})\times \mathbf{S}$, i.e. the \textit{total number} of $B$ sites in clusters of size $\mathbf{S}$ are plotted against $\mathbf{S}$ in double logarithmic scale. For each $\mathrm{p}$, statistics were collected over $100$ randomly generated binary matrix configurations. }
\label{Size}
\end{figure}

It is seen that in \ref{N(S)}, i.e. for $p=0.5$ the number of $B$ clusters is non-zero continuously over a wide range of cluster sizes. However for $p=0.75$, clusters are divided into two groups, a small group of small-sized clusters and a large group of very large sized clusters. The two groups are separated by a wide white gap occupied by  no $B$ cluster.

The same data can be presented on a double logarithmic scale, and with slight modifications as well,  to bring out some more features clearly, at higher values of $p$. This is done in figure \ref{SN(S)}. 

\begin{figure}
\centering
\begin{tikzpicture}
\node (img){\includegraphics[width=0.5\textwidth, trim=0cm 0cm 0cm 0cm, clip]{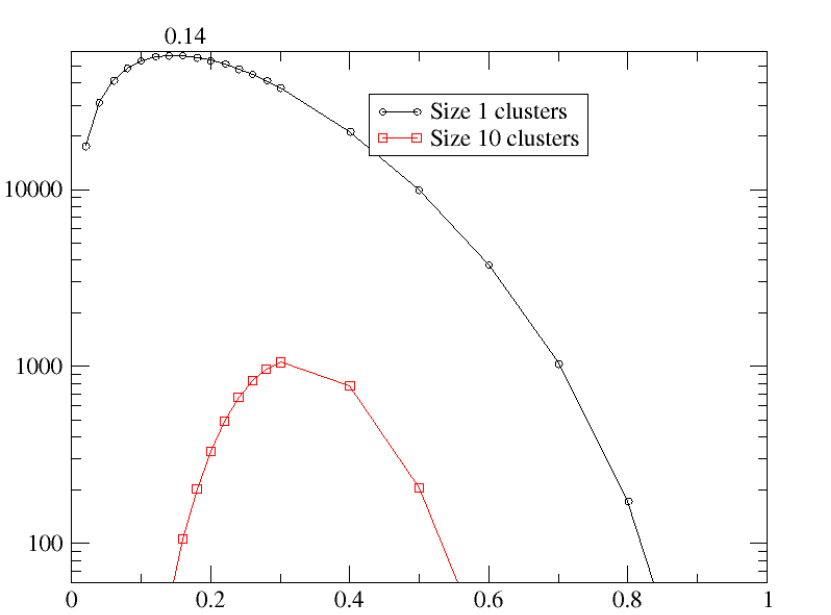}};
\node[below=of img, node distance=0cm, xshift=0cm, yshift=1cm,font=\color{black}] {$\mathbf{p}$};
\node[left=of img, node distance=0cm, rotate=90, anchor=center,yshift=-0.7cm,xshift=-0.5cm, font=\color{black}] {Number of Clusters};
\end{tikzpicture}
\caption{This plot shows how the number of clusters of two specific sizes 1 and 10 vary with $p$.}
\label{inset}
\end{figure}

Figure \ref{N(S)} shows the number of $B$ clusters $N(S)$ of size $S$ as function of $S$ and figure \ref{SN(S)} shows$N(S)\times S$ i.e. the \textit{total number} of $B$ sites in clusters of size $S$. From both figures it is evident that for $p < 0.5$ clusters of sizes varying continuously from $1$ to a specific value which increases with $p$ occur. However when $p$ reaches $0.5$ clusters of nearly all sizes are present, this is a signature of the percolation threshold. This appears very prominently as a broad continuous patch of colour in both figures \ref{N(S)} and \ref{SN(S)}.  As soon as the threshold is crossed clusters are divided into two highly discrete groups, a few very small clusters and a few very large clusters with no clusters of intermediate size. Ultimately at $p=1$, there is only one $B$ cluster covering  the whole system. Figures \ref{Size} and \ref{inset} results also corroborate this analysis.

As an example of how the number of clusters of a definite size varies with $p$ we show in figure \ref{inset} the variation of the number of  $B$ clusters of sizes 1 and 10. As $p$ starts to increase from 0, initially of course clusters of size 1 are most numerous, their number increases, reaches a peak and then starts to fall, ultimately reaching zero. In the meantime, larger clusters begin to form, the number of size 1 clusters is however never overtaken by clusters of larger size. The numerical results for the number of size 10 clusters is shown here for comparison. Interestingly, this is true in general for clusters of any size larger than 1 and is proved mathematically in appendix \ref{Appendix B}.

\FloatBarrier

\clearpage

\section{Discussion}\label{Sec 3}
\subsection{Euler Number Variation with Diagonal Connection Probability $q$}\label{Sec 3.1}

The Euler number graph (figure \ref{EulerNumber}) varies in an interesting manner as $q$ gradually increases from $0$ to $1$. 

\begin{itemize}
\item
When $q=0$, the connection probability of any two diagonally placed black pixels is $0$, whereas the connection probability of any two diagonally placed white pixel is $1$. Intuitively speaking, in such a situation, white clusters would have greater joining tendency as compared to black clusters.  Thus, at $p=0.5$, number of black clusters should exceed the number of white clusters, which in turn implies that $\chi(0.5)>0$. Also, clearly $\chi(p)>0 \ \forall \ p<0.5$. $\chi(p)$ would become negative beyond some value of $p$, say $p_0$, which is greater than $0.5$. $p_0$ may be estimated by considering a large number of system configurations at $q=0$. However, the value isn't deterministic.
\item
When $q=0.5$, the connection probability of any two diagonally placed black pixels is same as the connection probability of any two diagonally placed white pixels i.e. $0.5$. In this case, logically, the mean value of $p_0$ considering a large number of system configurations should be $0.5$.
\item
When $q=1$, the connection probability of any two diagonally placed black pixels is $1$, whereas the connection probability of any two diagonally placed white pixels is $0$. Thus, the black clusters would have greater tendency of joining compared to the white counterparts. At $p=0.5$, number of white clusters should exceed the number of black clusters, implying $\chi(0.5)<0$. We can also directly conclude that $\chi(p)<0 \ \forall \ p>0.5$ and that $\chi(p)$ should change from positive to negative, at some value of $p$ i.e. $p_0$ which should less than $0.5$. As mentioned earlier, the value of $p_0$ is not fixed for finite lattices, but may be estimated.
\end{itemize}

Interestingly, when $p_0$, the $B$ occupation probability where the number of black clusters and white clusters are equal, is plotted against $q$ (\ref{variation}), it is seen that the graph is approximately linear (even for a finite $1000\times 1000$ system). Linear regression on the data returns $p_0 = -0.2396 q + 0.6198$.

\begin{figure}
    \centering
    \begin{tikzpicture}
    \node (img) {\includegraphics[width=0.7\textwidth]{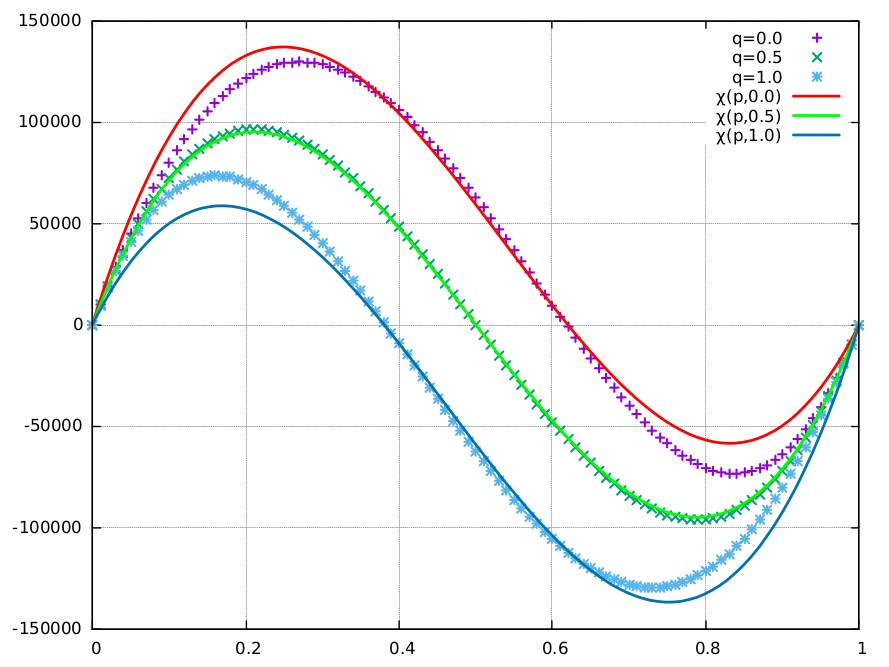}};
    \node[below=of img, node distance=0cm, yshift=1cm,xshift=0.8cm, font=\color{black}] {$\mathrm{\mathbf{p}}$};
    \node[left=of img, node distance=0cm, rotate=90, anchor=center,yshift=-0.7cm,xshift=0.2cm,font=\color{black}] { $\chi(\mathbf{p})$};
    \end{tikzpicture}
    \caption{Plot showing the cubic fit $\chi(\mathbf{p},\mathbf{q})$ for the Euler number $\chi=N_B-N_W$ statistics, collected over 100 random binary matrix configurations of size $1000\times 1000$, when $\mathbf{q}=0$, $\mathbf{q}=0.5$ and $\mathbf{q}=1$ respectively. The cubic approximation works almost perfectly when $\mathbf{q}=0.5$ but its accuracy decreases as $\mathbf{q}$ gradually shifts from $0.5$ towards either $0$ or $1$.}
    \label{CubicFitChi}
\end{figure}

Considering the appearance of the $\chi(p)$ graphs in figure \ref{EulerNumber} we try a cubic fit of the form $\chi(p)$ graphs, of the form $C(p-\alpha)(p-p_0)(p-\beta)=0 \label{cubic}$. Since the two end roots are nearly $0$ and $1$ respectively, we consider $\alpha=0$ and $\beta=1$. Applying a ``constant fit" on the data for $C$ we obtain $C = 1.97596 \times 10^{6}$. Thus, for practical (physical) systems we can approximate the Euler number $\chi$ as $\chi(p,q)=(1.97596 \times 10^6)(p - 0)(p - 1)(p - (-0.2396 q + 0.6198))$ (cf. Figure \ref{CubicFitChi}). The figure compares the simulation data for $q=0.0$, $0.5$ and $1.0$ represented by plus, cross and star symbols with respective data from solutions of equation (\ref{cubic}) represented by continuous red, green and blue lines.

\subsection{Variation of Spanning Cluster Percolation Threshold with Diagonal Connection Probability $q$}\label{Sec 3.2}

A classical definition of percolation phase transition in discrete percolation theory is based on the appearance of \textit{spanning clusters} \cite{grimmett_1999, stauffer}. Since we are concerned only with $2$ dimensional square lattices $\Lambda \in \mathbb{Z}^2$ with $V = L \times L$ sites, spanning clusters in this context are those clusters of occupied cells which either extend from the left border of the lattice to its right border, or from its bottom border to its top border. For infinite lattices, there exist a particular critical probability $P_c$, below which the probability of the existence of an infinite spanning cluster is $0$ but above which the probability of the existence of an infinite spanning cluster is $1$.  And indeed, $P_c$ is what we call the ``percolation threshold''. On a related note, the probability of the existence of a cluster spanning two given sides of a large box, or more generally, two arbitrary boundary segments, is sometimes referred to as the ``crossing probability''. Even for $L$ as small as $100$, the probability of the existence of a spanning cluster increases sharply from very close to zero to very close to one within a short range of values of $p$. This in itself hints at the underlying fact that finite large systems can be related to the $L\rightarrow \infty$ limit via the theory of ``finite size scaling". 

In figure \ref{ThresholdQ}, the offsets of the data points (w.r.t the Reference Line) for different $L$'s can be clearly seen to decrease with increasing $L$, and are hence expected to become zero in the infinite limit. 

\subsection{Size Distribution of Clusters}\label{Sec 3.3}

The nature of cluster sizes in the subcritical, critical and supercritical phases has always been an important topic of study in percolation theory.  We will discuss all the three phases one by one.

\begin{itemize}

\item \textbf{Subcritical Phase}: In the subcritical phase, $p<p_c$, the number of clusters of a certain size falls exponentially with the size. Further detailed discussion on this aspect are to be found in \cite{grimmett_1999,menshikov_mikhail,Aizenman1987,kesten_1982}.

\item \textbf{Critical Phase}: In the critical phase, where $p$ approaches $p_c$ sufficiently quickly as $n\to\infty$), the ratio between the largest cluster size $M_1$ and the second largest cluster size  $M_2$ follows a scaling law \cite{yong2015}. A detailed study of this feature may be planned in future for a range of $q$ values within the critical phase.

\item \textbf{Supercritical Phase}: In the supercritical phase, with $p$ tending to $1$ as $n\to\infty$, the largest $B$ cluster in an $n\times n$ system is of order approaching the system size. Moreover, the expectation value of the second largest cluster is sublinear in total number of sites \cite{Borgs2001}.

\end{itemize}

In our simulations the above characteristics appear to be present for all $q$, and we may conclude that the basic nature of cluster size distributions doesn't vary significantly with $q$ and $L$ (provided $L$ is sufficiently large, that is, at least $100$).

\clearpage

\section{Comparison with the Island-Mainland (IM) Transition Model}
\label{Sec 4}
\begin{figure}
\captionsetup[subfigure]{labelformat=empty}
\centering
\begin{tikzpicture}
\node (img){\subfloat[]{\label{main:a}\includegraphics[width=0.7\textwidth]{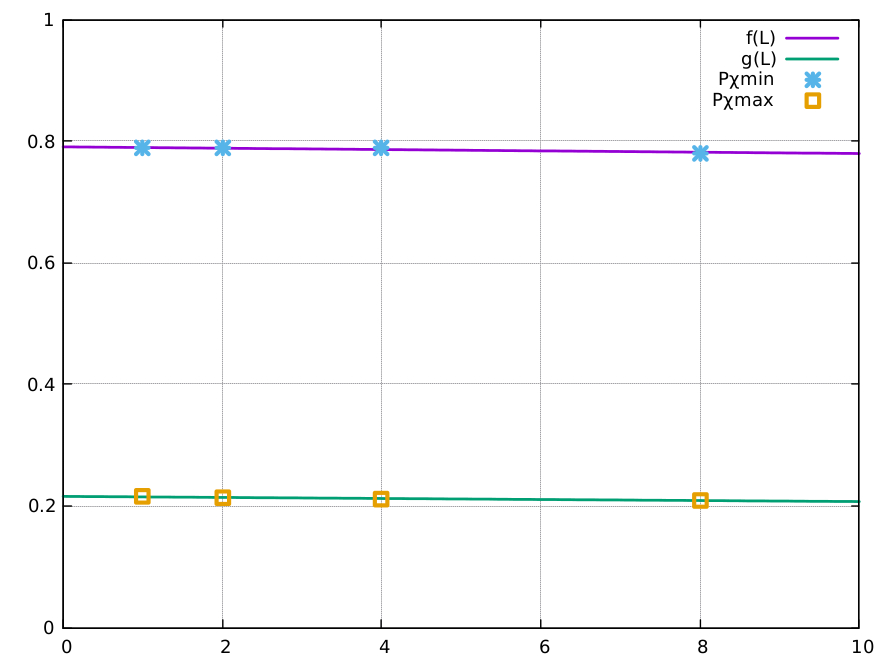}}};
\node[left=of img, node distance=-1cm, rotate=90, anchor=center,yshift=-0.8cm,xshift=0cm, font=\color{black}] {$\mathbf{p}$};
\node[below=of img, node distance=0cm, xshift=0cm, yshift=1cm,font=\color{black}] {$1000/\mathbf{L}$};
\end{tikzpicture}
\caption{Finite size-scaling on the maxima $\mathbf{p}_{\chi\text{max}}$ (squares) and minima $\mathbf{p}_{\chi\text{min}}$ (star symbols) of the Euler number curves for system sizes $\mathbf{L}=125, 250, 500$ and $1000$. Here $f(L) = 0.7906 - 0.0010 (1000/L)$ and $g(L) = 0.2303 - 0.0027 (1000/L)$.}
\label{minmax}
\end{figure}

In \cite{Khatun2017}, Khatun \textit{et al.} dealt with  random binary square lattices where cross connections were permitted. That is, say $d0$ is the probability of white cells being diagonally connected at crossover points, while $d1$ is the probability of black cells being diagonally connected at crossover points. They considered both $d0$ and $d1$ to be $1$. We successfully reproduced their simulations and verified the finite size-scaling limit (i.e. $L\to \infty$) of $P_{a1}$ and $P_{a2}$, where $P_{a1}$ is the value of $p$ at which the number of black clusters $N_B$ peaks and $P_{a2}$ is that value at which the number of white clusters $N_W$ peaks. The limiting values are named $p_{\text{max}B}$ and $p_{\text{max}W}$. We further performed finite size-scaling on the global maxima and minima of the Euler number curves $\chi(p)$, using the data generated for system sizes $L=125, 250, 500$ and $1000$ (averaged over $100$ iterations, as before). Let us call them $p_{\chi\text{max}}$ and $p_{\chi\text{min}}$ respectively. In the $L\to\infty$ limit, the values turn out to be $0.216 \pm 0.098\%$ and $0.791 \pm 0.196\%$, as illustrated in \ref{minmax}. 

In the same paper, $p_{c1}$ was defined to be that critical value of probability $p$, at which $N_W$ increases from $1$ to a value $>1$ i.e. the continuous white background breaks into two or more parts. Similarly, $p_{c2}$ was defined to the the critical value of $p$ at which the disjoint black clusters join to form a single large black cluster i.e. $N_B$ reduces to $1$. 

It was conjectured there, that $p_{c1}$ and $p_{c2}$ coincide with the maximum and minimum  of the Euler number curve - $p_{\chi\text{max}}$ and $p_{\chi\text{min}}$  respectively as $L\rightarrow \infty$. However here it is (see appendix \ref{Appendix A})  mathematically proved that as $L\rightarrow \infty$, $p_{c1} \rightarrow 1$ and $p_{c2} \rightarrow 0$ as $L\rightarrow \infty$, whereas from finite size scaling $p_{\chi\text{max}}$ and $p_{\chi\text{min}}$ tend respectively to the non-trivial values close to 0.2 and 0.8 respectively. So the quantities which survive finite size scaling are the two points where the derivative of $\chi(p)$ with respect to $p$ vanish or $$\Delta \chi(p)=0.$$
This implies that for a vanishingly small increase in $p$, say deposition of one black square, the change in the number of black clusters equals the change in the number of white clusters, or

$$\label{Ndiff}\Delta N_B(p_{a1}) = \Delta N_W(p_{a1})$$
and similarly for $p_{a2}$.

Adding a black site  can \textit{increase} $N_B$ when a new black square falls on a white site surrounded by eight others and can \textit{decrease} $N_B$ if the new black site unites 2 disjoint black clusters. The difference of these two quantities contributes to the left hand side of equation (\ref{Ndiff}). On the right hand side, $N_W$ can increase by a adding a black site, if it separates an existing white cluster into two disjoint clusters. Here $N_W$ can decrease if the new black site falls in an existing isolated black site.

In a real situation for example wetting/dewetting experiments, evaporation or condensation may not be random, but controlled by factors such as surface tension or adhesion. In such cases, these factors will control the probabilities of the above occurrences. Exploring such possibilities may be a useful application of the discussions presented.

Khatun et al.\cite{Khatun2017} described some experiments where the minimum in $\chi(p)$ was very close to the point where the background first broke up into disjoint clusters. We see here that for infinite systems this is not strictly true, but is more or less satisfied for real finite systems.

\clearpage

\section{Conclusion}
\label{Sec 5}

In this article we generate a strictly two-dimensional square lattice with a range of connection probabilities $q$ varying from 0 to 1, between second neighbour (diagonally placed) sites of same color (black or white). Nearest neighbour, i.e. edge-sharing sites of same color are always connected. This new feature ensures that black and white clusters are uniquely defined and not entangled or intertwined. The intertwining in the work of Feng \textit{et al.}\cite{feng} and the  quasi 3-dimensional nature in the work by Khatun \textit{et al.}\cite{Khatun2017} are thus avoided. Mertens and Ziff \cite{ziff} studied a special case of this problem with the Euler characteristic defined for the matching lattice. We have determined percolation thresholds for the whole range of $q$  and they are found to vary linearly. For the symmetric case with $q=0.5$ cluster size distributions and some other statistics have been determined.

We also point out an inconsistency in \cite{Khatun2017}. It was shown there that the maxima and minima for the Euler number $\chi_p$ converge to non-trivial values in the $L\rightarrow \infty$ limit and it was suggested that these values are identical to the values of $p$ where the white background broke up from a single connected cluster to more than one white and the single black cluster broke up into more than one black cluster. These points were named as IS(island) $\to$ MP(mixed phase) and MP $\to$ ML(mainland) transitions respectively. However, it is demonstrated here that these transitions do not survive finite size scaling as elaborated in Appendix \ref{Appendix A}, and are therefore not critical phase transitions. For real systems of finite size however, these observations work quite well.

An interesting difference is observed between the Euler number curve obtained in \cite{Khatun2017} with intertwined clusters and the Euler number curves in the present work. Khatun \textit{et al.} found inflection points in the Euler number curve corresponding to the values of the percolation thresholds. The Euler number curves in the present paper are smooth for all $q$ with no inflection points.

We may conclude by emphasizing the importance of the Euler number curve, in a percolating system under varied conditions of connection (such as varying $q$). Similar to the percolation threshold, the Euler number also survives finite size scaling. 

Problems worth further investigation in future may be (i) finding an explanation for the linearity of the $p_0$ vs. $q$ graph seen in figure \ref{variation} and (ii) working out a mathematical expression for the Euler number graphs for general values of $q$ and $p$ as obtained in figure \ref{EulerNumber}.

\section{Author Contributions}
\label{Sec 6}
SD and SS, undergraduate students at Jadavpur University, carried out the numerical computations and worked on mathematical analysis involved. The problem was conceived by ST and the project was carried out under the guidance of ST, TD and TK.
\section{Acknowledgement}
\label{Sec 7}
SD and SS acknowledge the support provided by the Condensed Matter Physics Research Centre, Jadavpur University during the period of the research project.

\bibliographystyle{plain}
\bibliography{custom}

\clearpage
\begin{appendices}
\section{Indeterministic transition probabilities in the ``Island Mainland'' problem}
\label{Appendix A}

We present a mathematical understanding of the nature of \textit{Island-Mainland} transitions \cite{Khatun2017}. Initially we will define a few terms which we will require subsequently. In a square lattice the probability of an element being occupied (alternatively $1$ or ``black") is considered to be $p$. $\mathrm{p_{c1}}$ is the supposed to the critical probability at which the number of white clusters increases from $1$ to any number greater than $1$ and, $\mathrm{p_{c2}}$ is the probability at which number of black clusters decreases from a number greater $1$ to $1$.

For clarity, let $\mathcal{C}_0, \mathcal{C}_1$ denote the clusters of 0's and 1's in the matrix (or graph), respectively. We are defining $\mathrm{p_{c1}}$ as $$ \mathcal{C}_{0} = \begin{cases} 1 & \qquad p < \mathrm{p_{c1}} \\> 1 & \qquad p > \mathrm{p_{c1}} \end{cases} $$ and $\mathrm{p_{c2}}$ as $$ \mathcal{C}_{1} = \begin{cases} 1 & \qquad p < \mathrm{p_{c2}} \\> 1  & \qquad p > \mathrm{p_{c2}} \end{cases} $$ for $N\to\infty$.
\noindent For a fixed $p \in [0,1]$ we can define
$$C_0(p) = \lim_{N \rightarrow \infty} \mathbf P_{p,N}[ |\mathcal C_0| > 1]$$ 
which is the limit of probability that there is more than one cluster of $0$s that is, more than one white cluster.
Let us define a critical probability $\mathrm{p_c}$ such that
$$
C_0(p) =
\begin{cases}
0 & \qquad p < \mathrm{p_c} \\
> 0 & \qquad p > \mathrm{p_c}
\end{cases}
$$
That is, when $p < \mathrm{p_c}$, for $N \rightarrow \infty$ the probability of having more than $1$ cluster is $0$. This would intuitively imply that there is at most $1$ cluster of $0$s in the limit. Conversely when $p > \mathrm{p_c}$ there is a positive chance (in the limit) of seeing more than 1 cluster.
\noindent For this definition of $\mathrm{p_c}$, it can be shown that $p_{c}$$= 0$.

To  verify this, supposing that $p>0$, the probability that a given $3 \times 3$ sub-matrix is given as :
$$ B = 
\left( \begin{matrix}
1 & 1 & 1 \\
1 & 0 & 1 \\
1 & 1 & 1
\end{matrix}
\right)$$
The probability of this configuration occurring is $q = p^8 (1-p) > 0$. Then suppose that we have  $N = 3n$, then the matrix can be seen as $n^2$ blocks like the above. Each of these $n^2$ blocks are independent, and have probability $q$ (which does not grow with $n$) of being of the form $B$. The number of the $n^2$ blocks which equals $B$ is given by a Binomial variable $\text{Bin}(n^2, q)$; in particular, the probability that more than two such blocks exist is
$$\mathbf{P}[ \text{Bin}(n^2, q) >= 2]$$ 
$$=1 - (1-q)^{n^2} - n^2q(1-q)^{n^2 - 1},$$

The probability of there being at least 2 clusters of $1$s is greater than the probability that at least two blocks like the above exist (since this is a special case of having two clusters), so that is

$$\lim_{n\rightarrow \infty} \mathbf P_{p,3n}[ |\mathcal C_1| > 1]$$ 
$$\geq \lim_{n \rightarrow \infty} \mathbf{P}[ \text{Bin}(n^2, q) >= 2]$$
$$ = \lim_{n \rightarrow \infty} 1 - (1-q)^{n^2} - n^2q(1-q)^{n^2 - 1}$$
$$= 1$$

That is, for any $p > 0$ we have $C(p) = 1$. Clearly $C(0) = 0$, and so it follows that $p_c =0$.

The decision to use $N = 3n$ is to make the proof a bit simpler. Further, with a bit more probabilistic machinery it can be argued via Kolmogorov's Zero-One law that in the limit the configuration $B$ appears infinitely often : which ensures that in fact for any $p > 0$ the expected number of clusters is infinite. 

Alternatively the same can be verified via simplified computation as well.If the matrix  is denoted $A_{i,j}$ with $1 \leq i,j \leq N$ consider just the $2 \times 2$ sub-matrix in the top left corner. If this takes the specific form

$$
C = 
\left( 
\begin{array}{cc}
0 & 1 \\
1 & 1
\end{array}
\right)
$$
and moreover if there is at least one more $0$ elsewhere in the matrix, then this would imply that we have two clusters of $1$s.

For any $N \geq 3$, and fixed $p > 0$ the probability that the top corner is equal to $C$ is given by $$(1-p) p^3$$
Of the remaining $N^2 - 4$ vertices, the number of $0$s to occur is distributed according to a $\text{Bin}(N^2 - 4, (1-p))$ variable, therefore the probability that there is at least one $0$ is
$$ 1 - \mathbf P[ \text{Bin}(N^2 - 4, (1-p) ) = 0] = 1 - p^{N^2 - 4}.$$
The probability of seeing the corner equal to $C$ and also there being at least one more $0$ is given by
$$q = p^3 (1-p) ( 1 - p^{N^2 - 4}).$$
Although this is a very special case of there being at least two clusters, but for any $p > 0$ the probability $q > 0$.

So we have that with probability $q$ a given sample will have this property. Whilst it is not certain how many samples are needed to see this particular event, on average one would expect to have to take $1/q$ samples.

For $N$ large, and $p$ small we can approximate $q \sim p^{3}$, so that $1/q \approx p^{-3}$. So for example when $p = 0.01$, we expect to need around $10^6$ samples to see this event.

This is an approximation to a very specific  example of having more than $1$ cluster. When taking into account the fact that there are four corners then the probability rises again, and approximately (assuming that each of the four possible corner events are independent, which is not the case) the probability at least one of the four occurring is
$$q' = 1 - (1-q)^4 \sim 1 - (1-p^3)^4,$$
which in turn means that on average it would take $1/q'$ samples before observing such a corner event. And noting that
$$ \frac1{q'} \sim \frac{1}{1 - (1-p^3)^4} \sim \frac1{4p^3} + \mathcal{O}(1)$$ we see that actually for $p = 0.01$ we need on the order of $250,000$ samples to see such a corner event.

From this, we can comment on the values of $\mathrm{p_{c1}}$ and $\mathrm{p_{c2}}$ as follows: as we see that $C_0(p)$ is a probability and starts rising in value as p rises from 0, the $\mathrm{p_{c1}}$ value for instance, may be 0 or some higher probability, with a greater probability of having a value nearer to 0. This may be verified with future simulations on the large size (infinite) model, to look for a trend of $\mathrm{p_{c1}}$ values approaching 0 and analogous to this $\mathrm{p_{c2}}$ values would approach(but span a probabilistic range near) 1. This shows that the values of $\mathrm{p_{c1}}$ and $\mathrm{p_{c2}}$ are not deterministic.

\section{Combinatorial reasoning for the descent in frequency of clusters with ascent in cluster size}
\label{Appendix B}

We present an idea of how the number frequency of small sized clusters in a large random matrix is always in descending order. This may explain our observations from computed results that for almost all except very high values of $p$ (that is as defined previously), the single cell clusters are most numerous. We look primarily at black clusters as occupied sites as before. For simplicity, we will consider that cells in the Moore neighborhood of any central cell and having the same color as that of the central cell, belong to the same ("occupied" or "unoccupied") cluster as of that central cell. Nevertheless, the primary conclusion of this discussion will apply to all diagonal connectivity patterns, up to second nearest neighbour.

This phenomenon is seen for $p < 1-\frac{1}{N^2}.$ on a random $N \times N$ matrix.
As an example,
for $p=1-\frac1{10^6}$ and a $1000 \times 1000$ grid one would expect on average one white cell  and $999,999$ black. The probabilities to see $0,1,2$ or $3$ cell clusters are about $36.8\%,36.8\%,18.4\%$ and $6\%.$\footnote{If an event has a probability $\frac{1}{M}$ and we do $M$ trials then the average number of hits is $1$ while the probability to get exactly one is almost exactly $\frac{1}{e} \approx 36.8\%$, and that is also the probability of getting no hits. For two hits it is $\frac{1}{2e}$. In general it is $\frac{1}{k!e}$ for $k$ small relative to $M$} So the largest cluster of black is $1000000$ or $999999$ a little over $\frac23$ of the time. However, if we are to make the grid $10^8 \times 10^8$ with that same $p$ value, we would get a definite descending order of frequencies.

The effect which causes this descending order trend is easily analyzed for a $1 \times N$ matrix. We consider $p$ as site occupation probability, and we record how long the cluster of occupied cells is each time we get such a cell. Letting $P_k$ be the probability that the next occupied cell we get will be the start of a cluster of length $k.$ It is easy to see that $P_k=p^{k-1}q$ so $P_{j+1}=pP_j < P_j$. This case is general and can be extended up to $N\to\infty$. If it were a finite $1 \times N$ rectangle then the chance that all the cells will end up black is $p_N=p^N$ so it is possible that $P_N > P_1 > P_2 > \cdots.$

Now we consider an $N \times N$ board. We assume $N$ is quite large and ignore effects at the corners and sides. From our observations of the simulation results, it can be said that for a large enough $p$ (around $p > 0.5$) there is usually one huge cluster and an assortment of smaller ones and there is an abrupt jump in the size of clusters formed around a value of $p$ near 0.5. This hints at the fact that the larger a partial cluster is (till a certain limit), the more likely it is to grow a bit more. This tends to spread out the larger sizes leaving no one occurring too often, and hence their frequencies are very low.

As a small case analysis: considering a cell not too near the edges. The probability that it is black in a cluster of size $1$ is $P_1=pq^8.$ There are $8$ ways it could be in a cluster of size $2.$ Half of them (shared side) require $10$ other squares to be white. The other four (shared corner) require $12$ white squares. So the probability to be in a cluster of size $2$ is $P_2=p^2(4q^{10}+4q^{12})$. Then, $P_2=4p(q^2+q^4)P_1.$ Solving for the maximum ratio we get that $P_2<0.9P_1$ with that bound occurring at about $p=0.27.$

The simplified analytical point of view can be seen as follows: We randomly assign the distinct weights $1,2,3\cdots$ up to $1000$ to the squares and then turn them white to black in that order. So we are gradually raising $p.$ We do this over a sufficiently large number of iterations. Usually there will gradually be a few  isolated one cell clusters far from each other. Eventually the first multi cell cluster will occur, probably of size $2$ but maybe $3$ or even $4$. But at that stage there are many single cell clusters. Eventually there will be more cells in multi-cell clusters than in single cell ones. But that distribution would have the number of clusters of sizes $1,2,3$ in a decreasing ratio, giving rise to our observed phenomenon.

The result we verify will definitely fail for $p=1$ and also, for an $N \times N$ board, if $p > 1-\frac{1}{N^2}.$ Then over $90\%$ of the time there are $0,1$ or $2$ white cells so for sure there is a single huge black cluster. There is, for a $1000 \times 1000$ board some critical probability $p_1$ above which the descending phenomenon fails. There is a probability of $pq^3$ at the four corners and of $pq^5$ at any one of the $3992$ other edge cells to be a single cell black cluster, making the 2D analysis somewhat tricky to extend from the 1D analysis as for the 2 row case, even above $p=0.5$ the frequency of 2 size clusters take over. But experimentally it is definitely verified that for larger sized $2D$ clusters at least for the first few natural numbers $n$, the number of clusters of size $n$ is greater than the number of clusters of size $n+1$. Around the site percolation threshold $p=0.407$ there seem to be some fluctuations, however the trend carries on in accordance with our findings, till around very near $p=1$, and the cluster sizes continue showing the above trend.
\end{appendices}
\end{document}